\documentclass[11pt,a4paper]{article}
\usepackage{jheppub}
\usepackage{graphicx,amsfonts,amssymb,amsmath}
\usepackage{bm}

\newcommand*{\ket}[1]{\mathopen{|}#1\mathclose{\rangle}}
\newcommand*{\bra}[1]{\mathopen{\langle}#1\mathclose{|}}

\begin{document}

\title{Holographic Geometries of one-dimensional gapped quantum systems from Tensor Network States.}
\author{Javier Molina-Vilaplana,}
\emailAdd{javi.molina@upct.es}
\affiliation{Department of Systems Engineering and Automation. Technical University of Cartagena \\ C/ Dr Fleming SN 30202, Cartagena, Spain.}

\abstract{We investigate a recent conjecture connecting 
the AdS/CFT correspondence and entanglement renormalization 
tensor network states (MERA). The proposal interprets the 
tensor connectivity of the MERA states associated to quantum 
many body systems at criticality, in terms of a dual holographic 
geometry which accounts for the qualitative aspects of the 
entanglement and the correlations in these systems. In this 
work, some generic features of the entanglement entropy and 
the two point functions in the ground state of one dimensional 
gapped systems are considered through a tensor network state. 
The tensor network is builded up as an hybrid composed by a 
finite number of MERA layers and a matrix product state (MPS) 
acting as a cap layer. Using the holographic formula for the 
entanglement entropy, here it is shown that an asymptotically 
AdS metric can be associated to the hybrid MERA-MPS state. 
The metric is defined by a function that manages the growth of 
the minimal surfaces near the capped region of the geometry. 
Namely, it is shown how the behaviour of the entanglement 
entropy and the two point correlators in the tensor network, 
remains consistent with a geometric computation which only 
depends on this function. From these observations, 
an explicit connection between the entanglement structure of 
the tensor network and the function which defines the geometry 
is provided.} 


\maketitle

\section{Introduction}\label{intro}
The best known concrete realization of the holographic principle 
is the AdS/CFT correspondence \cite{adscftbib}. This duality is 
generally formulated as the equivalence between the generating 
functionals $\mathcal{Z}$ of two very different theories: a 
classical theory of supergravity defined on a $D+1$ asymptotically 
Anti de Sitter spacetime and a QFT living on a $D$-dimensional 
spacetime that is the boundary of the AdS, i.e,
\begin{eqnarray}\label{adscft}
\mathcal{Z}_{QG}\left[ AdS_{D+1},\, \mathcal{J}(x,z) \right] \equiv \mathcal{Z}_{QFT}\left[\partial(AdS)_{D},\, J(x)\right] ~, 
\end{eqnarray} 

and more explicitly,
\begin{eqnarray}
e^{-S_{QG}[\mathcal{J}(x,z)]} \sim \left\langle e^{-\int J(x)\, \mathcal{O}(x)\, d^{D}x} \right\rangle_{\mathcal{J}(x,0)\equiv J(x)} ~,
\end{eqnarray}

 where $x$ are the spacetime dimensions of the QFT, $z$ is 
 the extra-dimension of the gravity theory on AdS, $\mathcal{O}$ 
 are operators of the quantum field theory on the boundary, and 
 $\mathcal{J}(x,z)$ are the classical fields of the gravitational 
 theory whose boundary values $\mathcal{J}(x,z)|_{z \to 0}$ act 
 as the sources  $J(x)$ for the correlation functions of the QFT.

Since the initial formulations of the correspondence, a huge 
amount of work has been carried out in order to generalize 
the AdS/CFT while providing numerous valuable models to 
study non-perturbative  effects in quantum field theory 
such as confinement and quantum phase transitions. However, 
still there is not a first principles derivation of the 
correspondence that allow us to explicitly construct a 
bulk gravity theory from a boundary QFT \cite{douglas11}. 
Nevertheless, it is widely accepted that the AdS/CFT is 
at heart a geometric formulation of the renormalization 
group (RG), in which the renormalization scale becomes 
the extra radial dimension $z$ and the beta functions 
of the boundary field theory are the saddle point 
equations of motion of the bulk gravity theory \cite{hologRG}. 

In recent years, ideas coming from quantum information 
have been especially relevant in establishing new numerical 
real space quantum renormalization group methods such as 
density matrix renormalization group (DMRG) \cite{white92}, 
and tensor network states (TNS) algorithms including matrix 
products states (MPS) \cite{Cirac07}, projected entangled-pair 
states (PEPS) \cite{Verstraete09}, multi-scale renormalization 
ansatz (MERA) \cite{vidalmera}, tensor renormalization group 
(TRG) \cite{Levin07}, and tensor-entanglement-filtering 
renormalization (TEFR) \cite{GuWen09}. These algorithms 
provide a set of variational \emph{ans\"atze} useful to 
characterize the low-energy (long distance) physics of 
quantum many-body systems. The TNS assumes a parameterization 
of a many-body wave-function by means of a collection of 
tensors connected into a network. The number of parameters 
required to specify these tensors is much smaller than the 
exponentially large dimension of the system's Hilbert space, 
allowing for the efficient representation of very large 
(and even infinite) systems. It has been recently proposed 
\cite{Evenbly11} that tensor network states can be broadly 
classified into two categories according to the geometry 
of the underlying networks. 

 In the first category, the network mimics the physical 
 geometry of the system, as specified by the pattern 
 of interactions in the Hamiltonian. The MPS ansatz for 
 one dimensional systems, which consists of a collection 
 of tensors connected into a chain and its generalization 
 for higher dimensional systems (PEPS), lie on this 
 \emph{physical geometry} category. 
 
 At variance, according to \cite{Evenbly11}, the collection 
 of tensors in the second category of TNS are connected so 
 as to parametrize the diferent length scales (or, equivalently, 
 energy scales) relevant to the description of the many-body 
 wave-function. These tensor networks organize the quantum 
 information contained in a state in terms of different 
 scales through a characteristic tensor connectivity 
 which spans an \emph{additional} dimension related 
 with the RG scale, as for instance in MERA.  This has 
 been used to define a generalized notion of holography 
 inspired by the AdS/CFT duality \cite{swingle09}. 
 There, the author has made the groundbreaking observation 
 that the tensor networks in MERA, happens to be a 
 realization of the AdS/CFT correspondence. In \cite{swingle09}, 
 the dual higher dimensional  holographic geometry of the 
 original system emerges when one realizes that the tensors 
 in MERA corresponding to a one dimensional quantum critical 
 point, are connected so as to reproduce a discrete version 
 of the  anti-de Sitter spacetime (AdS). After the initial 
 proposal, a substantial amount of work has appeared supporting 
 and extending the original idea \cite{Evenbly11, vanram09, 
 Molina11, vanram11, matsu1, soliton, matsu2, okunishi, 
 matsu3, ads_mera_tak, cMERA, swingle12, sptmera13}.

In terms of the AdS/CFT correspondence, the most salient 
feature of the holographic dual to a system with a 
characteristic length (energy) scale is the capping 
or truncation of the geometry for values of the radial 
coordinate close to this length scale. While the geometry 
remains approximately AdS for values of the radial 
coordinate smaller than the characteristic length 
scale, the capping of the geometry implies an infrared 
(IR) fixed point at a finite energy scale for the 
original CFT lying at the boundary of the AdS region, 
thus yielding the dual for a massive deformation of the theory. 

It is thus expected that the state at the IR scale will be a 
non entangled state if all the correlations vanish for it i.e 
there is no \emph{topological order}. However, it also might 
be the case of  nontrivially entangled IR states in some 
circumstances, which could be due to some underlying 
topological order. In \cite{soliton}, authors speculated, 
in terms of the AdS/MERA duality, about the structure of 
a tensor network that might suitably describe both possibilities.

The aim of this paper is to offer new insights on the 
connection between the structure of MERA states and 
their potential holographic descriptions by exploring 
some hints posed in \cite{soliton}. Here it is shown  
that the entanglement entropy and the two point 
functions in a type of hybrid tensor network state 
(composed by a finite number of MERA layers and an 
MPS acting as a cap layer) representing a one 
dimensional quantum many body system, remain 
consistent with a geometric computation of both 
quantities once a sensible emergent metric 
associated to the tensor network is provided. 
This metric corresponds to an asymptotically 
AdS geometry with an IR capping region characterized 
in terms of a function which keeps track of the growth 
of minimal surfaces near the capped region of the geometry. 
From these observations, an explicit connection 
between the entanglement structure of the tensor 
network and the function which defines the IR geometry 
is provided.

 The paper is organized as follows: In section two, 
the MPS and MERA tensor network states are briefly 
reviewed and the hybrid tensor network composed of 
a finite number of MERA layers and a capping MPS 
on the top layer is presented. Section three reviews, 
the AdS/MERA conjecture and its relation with the 
formula for the computation of the holographic entanglement 
entropy \cite{ryu}. In section four, the generic features 
of the entanglement entropy and the two point functions 
computed with the hybrid tensor network presented in 
section two are constrasted with the same quantities 
computed through holographic calculations in  
sensible-looking geometries that may be 
associated to the tensor network.  Finally, in 
section five, results are summarized and future 
research problems are exposed.

\section{Tensor Network States for 1D systems}\label{TNS:sec}
Tensor network states (TNS) constitute a new class of numerical 
methods  that aim to efficiently describe ground states 
of strongly correlated quantum systems. TNS implement 
techniques drawed from renormalization group methods 
and the knowledge about the structure of entanglement 
in the ground  states of quantum many body systems. 
In this section we briefly review some issues concerning 
the two classes of tensor network states (MPS and MERA) 
used in this paper in order to offer further evidence 
on the connection between AdS/CFT and MERA.

\subsection{Matrix Product States}\label{MPS:Sec}
Let us first to consider a 1-D translationally invariant 
quantum many body system composed by $N$ sites of dimension 
$d$. Its ground state can be expressed as 
\begin{eqnarray}\label{mps}
\ket{\Psi} = \sum_{s_1,s_2,\cdots, s_N\, = 1}^{d}  
 \; {\mathcal T}_{s_1,s_2,\cdots, s_N} 
\ket{s_1,s_2,\cdots, s_N}\;, \label{state}
 \end{eqnarray}

 where for $j\in\{1,\cdots,N\}$ and $s_j \in\{ 1, \cdots, d\}$, 
 the vectors  $\ket{s_j}\in{\cal H}_d$ form the computational 
 basis of the $j$-th system site and where the 
 type - $\mbox{\tiny{$\left(\begin{array}{c} 0 \\ N \end{array}\right)$}}$ 
 tensor  ${\cal T}_{s_1,s_2,\cdots, s_N} = \bra{s_1,s_2,\cdots, s_N} \Psi \rangle $ 
 represents the associated  probability amplitudes. 
 The MPS representation of this state is given by the ansatz \cite{Cirac07},
\begin{eqnarray}\label{mps_cof}
\mathcal{T}_{s_1,s_2,\cdots, s_N} = {\rm Tr}\, \left( A^{\it{s_1}} \,
 A^{\it{s_2}}\,\cdots A^{\it{s_N}}\, \right) ~,\end{eqnarray}

where $A^{\it{s_j}}$ are matrices of dimension $\mathcal{W}$ 
with elements $A^{s_j}_{\alpha, \beta}$  whose values are 
determined by means of a variational search procedure. 
Away from the critical point, a finite $\mathcal{W}$ 
suffices to exactly describe the ground state. Essentially, 
the  so called bond dimension  $\mathcal{W}$, controls 
how much entanglement between adjacent sites is retained 
during the variational procedure. Treating the matrices 
$A^{\it{s_j}}$ as  $\mbox{\tiny{$\left(\begin{array}{c} 1 \\ 2 \end{array}\right)$}}$ 
tensors (one extra index is due to the physical site), 
the MPS ansatz can be pictorially  described as in figure 1. 

\begin{figure}[t]
 \centering
 \includegraphics[width=0.75\textwidth]{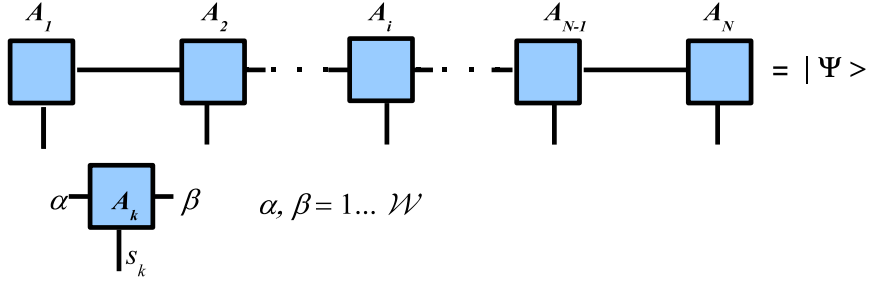}
 \caption{Pictorial representation of the MPS \emph{ansatz}. 
 The matrices $A^{\it{s_k}}$ are three indices tensors 
 (excepting the first $A^{\it{s_1}}$ and last $A^{\it{s_N}}$ 
 tensors, which only have two indices due to open boundary 
 conditions), with $\it{s_k}$ related with the physical site 
 $k$ and $\alpha$ and $\beta$ as the indices of the correlation 
 space of dimension $\mathcal{W}$ linking the two neighboring 
 matrices $A^{\it{s_{k-1}}}$ and $A^{\it{s_{k+1}}}$. In the figure, 
 connecting two adjacent tensors through a given link is equivalent 
 to contracting the product of these two tensors through the 
 corresponding index. Applying this rule to all the matrices 
 of the picture results in equation (\ref{mps_cof}).} 
\end{figure} 

Regarding the computation of observables with the MPS ansatz, 
given the state (\ref{mps}) with coefficients (\ref{mps_cof}), 
the expectation value of an operator $\Theta = \mathcal{O}_1 \otimes \mathcal{O}_2  \otimes\,
 \cdots \otimes \mathcal{O}_N $ which is the tensor product 
 of local operators $\mathcal{O}_j$ for each site $j$, 
 can be expressed as (Fig.2),
\begin{eqnarray}\label{ExVal_mps}
\bra{\Psi} \Theta \ket{\Psi} = 
\rm{Tr} \left( \mathbb{E}_{\mathcal{O}_1}\,\mathbb{E}_{\mathcal{O}_2}\, \cdots\, \mathbb{E}_{\mathcal{O}_N}\, \right)~.
\end{eqnarray}

 The \emph{transfer matrices} $\mathbb{E}_{\mathcal{O}_j}$ 
 are defined by,
\begin{eqnarray}\label{trans_Matrix}
\mathbb{E}_{\mathcal{O}_j}= \sum_{s_j, s^{'}_j = 1}^{d} \bra{s^{'}_j} \mathcal{O}_j \ket{s_j}\,
 \left( A^{s_j} \otimes \bar{A}^{s^{'}_j} \right)~,
\end{eqnarray}

and $\bar{A}^{s^{'}_j}$ is the hermitian conjugate 
of $ A^{s_j}$. As a result, as posed in equation 
(\ref{ExVal_mps}), the computation of expectation 
values with MPS, may be viewed as the total 
contraction of a 1-dimensional array of tensors 
($\mathbb{E}_j$).

\begin{figure}[t]
  \centering
  \includegraphics[width=0.65\textwidth]{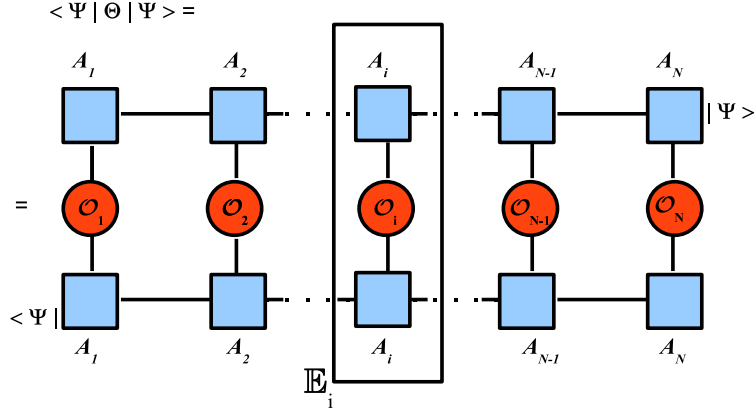}
  \caption{Pictorial representation of an expectation 
  value computation through the MPS \emph{ansatz} given 
  by equations (\ref{ExVal_mps}) and (\ref{trans_Matrix}).}
\end{figure}

It is well known that an MPS with a finite bond 
dimension $\mathcal{W}$, faithfully represents 
the ground state of a 1D gapped system and supports 
the exponential decay of correlations expected for 
these models. Considering a translational invariant 
matrix product state $\ket{\Psi}$, homogeneously 
defined  by the matrices $A_s$, the two point 
correlation function of two  local observables 
$\Theta_{\alpha}$ and $\Theta_{\beta}$, with 
support on sites $s_i$ and $s_j$ separated an 
arbitrary distance $\ell+1$ is defined by,
\begin{eqnarray}\label{MPS:2pcorr}
 \mathfrak{C}_{\alpha\, \beta}^{[\ell+1]} \equiv \bra{\Psi} \Theta_{\alpha}(s_i) \, \Theta_{\beta}(s_j) \ket{\Psi}={\rm Tr} \left(\mathbb{E}_{\mathbf{1}}^{[s_1]} \cdots  \mathbb{E}_{\Theta_{\alpha}}^{[s_i]} \cdots \mathbb{E}_{\Theta_{\beta}}^{[s_j]} \cdots \mathbb{E}_{\mathbf{1}}^{[N]}\right)~.   
\end{eqnarray}

 Following \cite{scholl10}, the correlator 
 in the limit when $N \to \infty$ can be written as:
\begin{eqnarray} \label{MPS:2point}
   \mathfrak{C}_{\alpha\, \beta}^{[\ell+1]}  =
   \sum_{\nu \geq 2}\, c_{\nu}\, (\lambda_{\nu})^{\ell}~,
\end{eqnarray}
where $\lambda_{\nu \geq 2}$ are the eigenvalues 
of $\mathbb{E}_{\mathbf{1}}$ for which it holds 
that $\vert\lambda_{\nu \geq 2}\vert < 1$. The 
coefficients $c_{\nu}$ are given by,
\begin{equation}\label{mps:coeff}
c_{\nu}= \bra{{\rm L}_1}\mathbb{E}_{\Theta_{\alpha}}^{[s_i]}\ket{{\rm R}_{\nu}}\, \bra{{\rm L_{\nu}}} \mathbb{E}_{\Theta_{\beta}}^{[s_j]}\ket{{\rm R}_1}~,
\end{equation}
 where $\ket{{\rm R_{\nu}}}$ and $\bra{{\rm L_{\nu}}}$ 
 are the right and left eigenvectors of 
 $\mathbb{E}_{\mathbf{1}}$ for the eigenvalues $\vert\lambda_{\nu}\vert < 1$ 
 and $\ket{{\rm R_{1}}}$ and $\bra{{\rm L_{1}}}$ are the right 
 and left eigenvectors of 
 $\mathbb{E}_{\mathbf{1}}$ for the eigenvalue $\lambda_1 = 1$.

 Since every $\vert\lambda_{\nu \geq 2}\vert < 1$, 
 then the correlator $\mathfrak{C}_{\alpha\, \beta}^{[\ell+1]}$ 
 decay exponentially because it is possible to write its 
 leading behaviour as a superposition of exponentials with 
 decay lengths defined by,
 \begin{equation}\label{mps:corrlength}
  \xi_{\nu}=-\frac{1}{\log \vert \lambda_{\nu}\vert}~.
  \end{equation}
  
 This implies that the MPS gapped nature dominates when 
 probing long range correlations, which are ruled by the eigenvalues 
 $\lambda_{\nu}$ of the identity transfer matrix $\mathbb{E}_{\mathbf{1}}$.
  
  Furthermore, if one computes the entanglement entropy 
  of a subsystem $A$ within an 1D-MPS state, bipartitioning 
  the system into two subsystems $A$ and $B$, and then 
  taking the partial trace over the sites on $B$, it is 
  easy to convince oneself that,
\begin{eqnarray}\label{EE:mps}
S_A \leq 2\, \log \mathcal{W}~,
\end{eqnarray}
 
 as one realizes that the bipartition in a 1 dimensional 
 MPS only involves two bond indices, which maximal 
 contribution to $S_A$ is given by $\log \mathcal{W}$. 
 Thus, $S_A$ does not scale with the size of the 
 subsystem $A$ despite it can be made arbitrarily 
 large by increasing $\mathcal{W}$. As the entanglement 
 entropy in 1D critical systems scales proportional 
 to $\log \ell$, where $\ell$ is the size of the 
 subsystem $A$ \cite{hlw94,calcar04}, then, from 
 the bound (\ref{EE:mps}) one clearly sees that an 
 MPS does not support the large amount of entanglement 
 required in quantum critical systems.

\subsection{Entanglement renormalization tensor network states. MERA}\label{MERA:Sec}
The MERA representation of the 1-$D$ state (\ref{mps}) 
assumes a decomposition of $\mathcal{T}_{s_1,s_2,\cdots, s_N}$ 
in terms of a collection of smaller, finite size tensors, which 
differently from the  linear MPS structure, are  organized in  
a  two-dimensional layered graph. The sites of the graph 
represent tensors. They are divided in two groups: the  
type $\chi\,- \mbox{\tiny{$\left(\begin{array}{c} 2 \\2  \end{array}\right)$}}$ 
tensors with elements ${\chi}^{u_1,u_2}_{s_1,s_2}$  
called \emph{disentanglers} and  the type $\Lambda\, - \mbox{\tiny{$\left(\begin{array}{c} 1 \\2  \end{array}\right)$}}$  tensors with elements 
$\Lambda^{u_1}_{s_1,s_2}$ called \emph{isometries} (Fig. 3). 
For quantum critical systems, a MERA tensor network posses 
a characteristic scale invariant structure, i.e, a unique 
$\chi$ and a unique $\Lambda$ precisely define the MERA graph.  

The MERA representation of (\ref{mps}) implements an efficient 
real space renormalization group procedure through a tensor 
network organized in different layers labelled by $\tau$. 
Each layer of MERA defines a RG transformation: prior to the 
renormalization of a block of typically two sites located at 
layer $\tau$ into a single site by means of a $\Lambda^{\tau}$-type 
tensor, short range entanglement between the sites is removed by 
means of the {\em disentangler} $\chi^{\tau}$. In other words, 
the MERA tensor networks, as a consequence of the locality of 
physical interactions in the ground state of  many-body systems, 
at each layer, decouple the relevant low energy degrees of freedom 
from the high energy ones, which are then safely removed, by 
unitarily transforming with \emph{disentanglers} small regions 
of space. The MERA coarse-graining transformation induces an 
RG map that can be applied arbitrarily many times keeping 
constant the computational cost for obtaining consecutive 
effective theories for the original system which are also 
labelled by $\tau$.

For 1D systems, the MERA computation of an observable $\Theta$ 
requires the contraction of a 2D tensor network, and this 
may be efficiently done by the  isometricity requirements 
for $\chi$ and $\Lambda$ tensors (see figure 4). Indeed, 
in an scale invariant MERA, contracting an entire layer 
of MERA and its conjugate, maps the original expectation 
value problem into a new one i.e,
\begin{eqnarray}
\bra{\Psi} \Theta \ket{\Psi} = \bra{\Psi^{'}} \mathcal{S}\left(\Theta\right) \ket{\Psi^{'}}= \bra{\Psi^{'}} \Theta^{[1]} \ket{\Psi^{'}}~,
\end{eqnarray}
where the quantum state $\ket{\Psi^{'}}$ is the original 
MERA state $\ket{\Psi}$ after the full contraction of 
the tensors at the bottom layer with their conjugates,
 the observable $\Theta^{[1]}=\mathcal{S}\left(\Theta\right)$ 
 is the \emph{effective} tensor shown inside the rectangle 
 in figure 4 (right), and $\mathcal{S}$ is the 
 \emph{scaling superoperator} \footnote{A superoperator 
 is a linear operator acting on a linear space of linear 
 operators. The prefix super- have no connection to supersymmetry 
 or superalgebras.} of the observable $\Theta$. The total contraction 
 of the tensor network proceeds by mapping the effective 
 observable in a sequential way as,
 
\begin{figure}[t]
  \centering
  \includegraphics[width=0.75\textwidth]{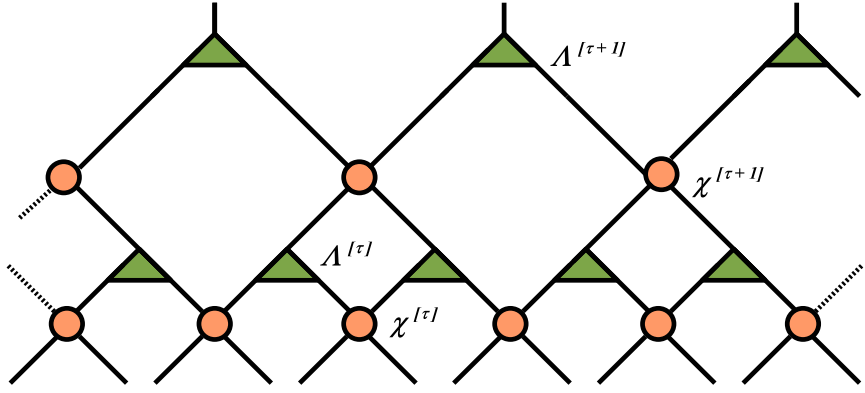}
  \caption{Pictorial representation of two layers 
  ($\tau$ and $\tau + 1$) of an scale invariant MERA 
  tensor network. Three legged triangles represent 
  isometries $\Lambda$ and four-legged circles represent 
  disentanglers $\chi$. At each layer $\tau$, the 
  tensors are chosen so as to fulfill $\chi^{\dagger}\, \chi=\mathbb{I}$ 
  and $\Lambda^{\dagger}\, \Lambda=\mathbb{I}$.}
\end{figure}

\begin{eqnarray} 
 \Theta^{[\tau]} \to \Theta^{[\tau+1]} = \mathcal{S}(\Theta^{[\tau]})~,
\end{eqnarray}
until the top of the tensor network $\ket{\mathcal{C}}$ 
is reached, for which it holds that
$\bra{\Psi} \Theta \ket{\Psi} =  \bra{\mathcal{C}} \Theta^{[h]} \ket{\mathcal{C}}$, 
with $h = \log_2 N$ and $N$ being the total size of the system. 

One of the most salient features of an scale invariant 
MERA tensor network is that it naturally supports 
the well known power-law decaying of correlation 
functions in systems at the quantum critical point 
\cite{meracft}. To show this, it is convenient to 
define the fixed point \emph{one site scaling superoperator}
$\mathcal{S}\equiv \mathcal{S}_{(1)}$ as, 
\begin{eqnarray}
\mathcal{S}_{(1)}(\circ) = \sum_{\alpha}\, \mu_{\alpha}\, \Phi_{\alpha}\,  \rm{Tr}(\Phi_{\alpha}\, \circ)~, 
\label{spectral1}
\end{eqnarray}
 
where the scaling dimensions of the scaling 
operators $\Phi_{\alpha}$ of the theory are 
$\Delta_{\alpha}\equiv -\log_2 \mu_{\alpha}$ 
and $\rm{Tr}(\Phi_{\alpha}\, \Phi_{\beta}) = \delta_{\alpha\beta}$. 
Let us consider the correlator of two scaling operators 
$\Phi_{\alpha}(s_i)$ and $\Phi_{\beta}(s_j)$ initially 
located at sites $s_i$ and $s_j$ of the original lattice, 
\begin{eqnarray}
\mathfrak{C}_{\alpha\beta} = \bra{\Psi} \Phi_{\alpha}(s_i)\,  \Phi_{\beta}(s_j)\ket{\Psi} = \left\langle \Phi_{\alpha}(s_i)\,  \Phi_{\beta}(s_j)\right\rangle~. 
\label{C2}
\end{eqnarray}

 After $\tau^{*} = \log_2 \vert s_i -s_j \vert$ 
 contractions of the type described above, the 
 sites supporting the original $\Phi_{\alpha}$ 
 and $\Phi_{\beta}$ become first neighbors 
 \cite{meracft} and each iteration contributes 
 a factor $\mu_{\alpha}\, \mu_{\beta}$ giving,
\begin{eqnarray}
\mathfrak{C}_{\alpha\beta} = 
\bra{\mathcal{C}^{*}} \left( \mathcal{S}_{(1)}(\Phi_{\alpha}(s_i)\right)^{\tau^{*}} \left(  \mathcal{S}_{(1)}(\Phi_{\beta}(s_j) \right)^{\tau^{*}} \ket{\mathcal{C}^{*}}=(\mu_{\alpha} \mu_{\beta})^{\tau^{*}}\bra{\mathcal{C}^{*}}\Phi_{\alpha}(0)\,  \Phi_{\beta}(1)\ket{\mathcal{C}^{*}}~,\label{scaling3}
\end{eqnarray}

 where the state $\ket{\mathcal{C}^{*}}$ is the 
 result of $\tau^{*}$ contractions of the original 
 MERA state with its conjugate. Defining $\mathfrak{C}_{\alpha\beta}^{\tau^{*}}\equiv\bra{\mathcal{C}^{*}}\Phi_{\alpha}(0)\,  \Phi_{\beta}(1)\ket{\mathcal{C}^{*}}$, noticing that 
 $\Delta_{\alpha,\,  \beta}\equiv -\log_2 \mu_{\alpha, \beta}$ 
 and recalling the identity $a^{\log b}= b^{\log a}$, 
 the two point function (\ref{C2}) can be written as,
\begin{eqnarray}\label{mera:scaling}
\left\langle \Phi_{\alpha}(s_i)\,  \Phi_{\beta}(s_j)\right\rangle  = \frac{\mathfrak{C}_{\alpha\beta}^{\tau^{*}}}{\vert s_i -s_j \vert^{\eta}}~,
 \label{merascaling}
\end{eqnarray}

with $\eta=(\Delta_{\alpha} + \Delta_{\beta})\delta_{\alpha\beta}=2\Delta_{\alpha}$. 

\begin{figure}[t]
  \centering
  \includegraphics[width=0.75\textwidth]{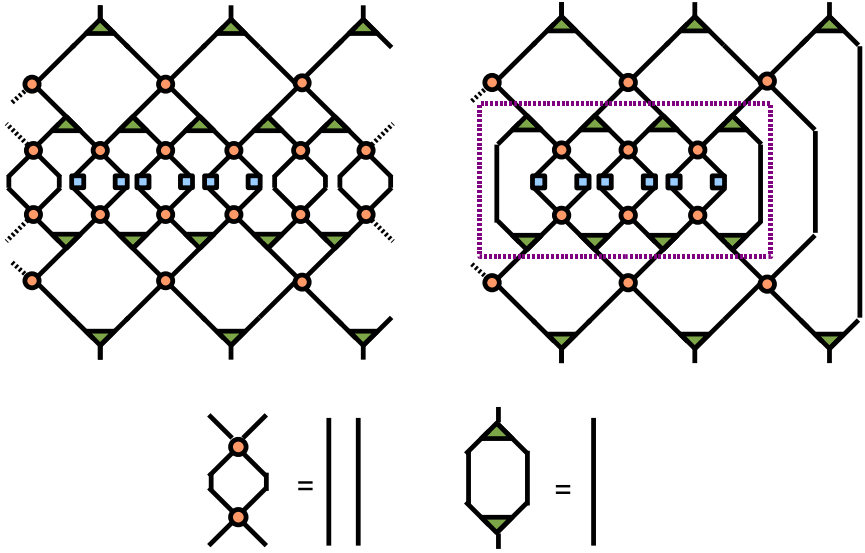}
  \caption{To compute the expectation value of an observable 
  $\Theta$ (blue squares) supportted on an interval of sites, 
$\{s_1 .. s_k \}$  with a 1D MERA,  the first layer of 
disentanglers $\chi^{[1]}$, for which, by definition, holds 
$\chi^{\dagger} \chi = \mathbb{I}$ is initially considered.
Those $\chi_{r_1\, r_2}^{u_1,\, u_2}$ whose (both) physical 
indices do not belong to the support interval $\{s_1 .. s_k \}$, 
are trivially contracted with their adjoints. However, this 
contraction is not trivial for those disentanglers directly 
connected to the support. The same argument holds for the 
isometries $\Lambda$ belonging to the first layer of the MERA 
state.}
\end{figure}

\subsection{Hybrid MERA-MPS networks}\label{Hybrid:Sec}
When considering the ground state $\ket{\Psi}$ of gapped 
1D-Hamiltonian $\mathcal{H}$, the correlations decay 
exponentially at large distances due to the characteristic 
correlation length $\xi$, while typically keep power-law 
decaying for distances smaller than $\xi$. To emulate 
this behavior with a tensor network state, it is necessary 
to have an ansatz which suitably reproduces these features. 
A potential candidate for this, is a tensor network state 
with an MPS at the top of a finite number ($\tau_0 \sim \log \xi)$ 
of MERA layers \cite{Evenbly11}. Here it is assumed that these 
$\tau_0$ layers correspond to those in the scale invariant 
MERA describing a neighbouring critical point while the 
capping MPS accurately describes the behaviour of the 
long distance correlations in the gapped phase. 

 In the MERA representation of the ground state of a gapped 
 Hamiltonian, after $\tau_{0}\approx \log \xi$ renormalization 
 steps, the original ground state $\ket{\Psi_0}$ of the system 
 has flowed into a state $\ket{\Omega}$ that can be well 
 approximated by a product state with no entanglement between 
 the different coarse grained lattice sites. This state describes 
 the ground state at a fixed-point of the RG flow corresponding 
 to a gapped phase without topological order. Nevertheless, 
 the finite layered MERA may also be combined with another 
 tensor network (an MPS in figure 5) in order to represent 
 the ground states of gapped systems that flow towards non 
 trivially entangled IR fixed point ground states $\ket{\Phi}$.

 \begin{figure}[t]
  \centering
  \includegraphics[width=0.75\textwidth]{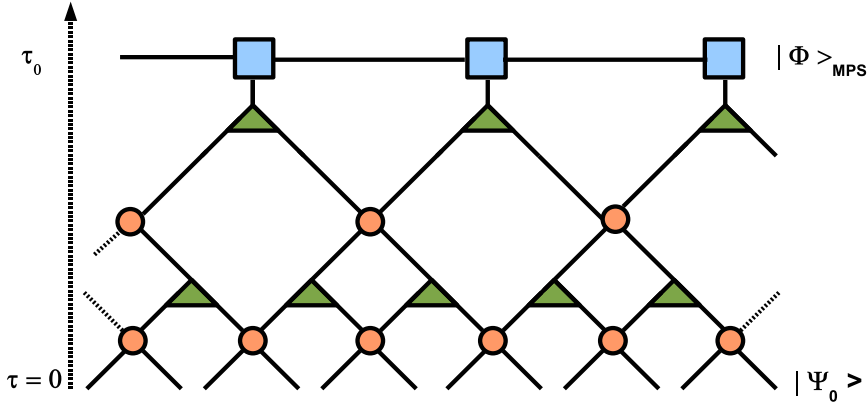}
  \caption{Pictorial description of an Hybrid 
  MPS-MERA tensor network state with a fixed number 
  $\tau_0$ of MERA layers and a top tensor given by 
  an MPS. The original ground state of the system 
  $\ket{\Psi_0}$ lies at the first layer of the MERA 
  \emph{curtain} while the ground state at the IR 
  fixed point is given by the MPS tensor capping 
  the MERA geometry.}
\end{figure}

The MPS tensor at the top layer of this hybrid MPS-MERA state, 
accounts for the gapped character and non trivial long range 
entanglement of the IR fixed point in a natural way, i.e, it 
naturally arranges the exponential decay of the two point 
correlations at distances larger than $\xi$, while the MERA 
'\emph{curtain}' of $\tau_0$ layers implements the power-law 
decaying behaviour of correlators when the distance between 
the physical sites is smaller than $\xi$. The case of the 
trivial non-entangled product IR state $\ket{\Omega}$ is 
represented by an MPS with bond dimension $\mathcal{W}=1$.

In the MPS-MERA tensor network, the total system 
size $N$ is given by,
\begin{eqnarray}
N = 2^{\tau_0}\, m~,
\end{eqnarray}

where $m$, is the number of \emph{effective} MPS 
sites located at the top of the $\tau_0$-th 
layer of the binary MERA '\emph{curtain}' hanging 
from the MPS tensor. Regarding the entanglement 
that may be supported by these type of tensor 
states, as each "site" in the top MPS represents 
a cluster of $\sim 2^{\tau_0}$ coarse grained strongly 
correlated sites of the original lattice, then the 
entanglement between different sites in the MPS top 
tensor represents the short-range entanglement between 
these clusters. If the MPS top tensor is not the trivial 
product state, its short-range entanglement then describes 
a particular pattern of long-range entanglement between 
distant sites of the original lattice. For finite $\mathcal{W}>1$, 
this pattern of long-range entanglement, which relates with 
some kind of topological order \cite{wen11, topord1}, clerly differs 
from the characteristic pattern appearing in quantum critical 
systems described by a non-capped scale invariant MERA network, 
while providing a non zero contribution to the total entanglement 
entropy of a region $A$ in the original lattice. 

In the hybrid MERA-MPS tensor network, the local unitary 
transformations carried out by the disentanglers of the MERA 
"\emph{curtain}" only remove the short range entanglement, 
between the neighboring coarse grained sites lying on the 
\emph{curtain}. 
Nevertheless, the pattern of entanglement  
between distant sites of the original lattice encoded in the 
short range entanglement between clusters, is not removed 
by the MERA disentanglers. It is precisely this pattern of 
long range entanglement which  entails the non trivial 
topological order of the IR fixed point state represented 
by the MPS at the top of the MERA \emph{curtain}.
 
\section{The $\rm{AdS}$/MERA duality}\label{ads:mera}
In \cite{swingle09}, it was firstly observed that MERA
\cite{vidalmera} happens to be a realization of the AdS/CFT
correspondence \cite{adscftbib}. As pointed out in \cite{swingle09}, 
from the entanglement structure of a quantum critical many body system,
is possible to define a higher dimensional geometry in which, apart from 
the coordinates labelling the position and the time $t$, one may add a "radial" 
coordinate $z$ labelling the hierarchy of scales. Then, the higher dimensional 
 geometry emerging from MERA may be visualized by locating cells
around all the sites of the tensor network representing the
quantum state. The size of each cell is defined to be
proportional to the entanglement entropy of the site in the cell.
Through this procedure, a geometric dual picture of the tensor 
network arises quite naturally from the entanglement of the degrees of
freedom of the critical system lying on the boundary
\cite{vanram09, ads_mera_tak}.

The discrete geometry emerging at the critical point is a discrete
version of AdS. For a one-dimensional quantum critical system with 
a space coordinate labelled by $x$, the continuous isometry $w \to w +
\theta$, $x \to e^{\theta}x$ of the metric
\begin{eqnarray}\label{ads:mera1}
ds^2_{\rm AdS}\sim dw^2 + e^{-2w}\left( dx^2 - dt^2\right) ~,
\end{eqnarray}

is replaced by the MERA's discretized version, $w \to w + k$, $x
\to 2^{k}x$ or $x \to 3^{k}x$ depending on the binary or ternary
implementation of the renormalization algorithm \cite{vidalmera}.
Here, the extra direction $w$ in the AdS is identified 
with $\tau$ (the variable labelling the number of renormalization 
steps) in MERA. To be more concrete, the $(D + 1)$-dimensional 
AdS metric can be written in the form,
\begin{eqnarray}
ds^2_{\rm AdS} = \frac{L^{2}_{\rm AdS}}{z^{2}}\left(dz^2 + dx^2 -dt^{2}\right)~,
\label{AdS}
\end{eqnarray}

where, $L_{\rm AdS}$ is a constant called the AdS radius; it has
the dimension of a length and it is related with the curvature of
the AdS space. We have also made the change of variable $\tau=\log{z}$, where
$z$ is the standard radial coordinate of the Poincar\'e AdS. With this choice of the 
spacetime coordinates, the one dimensional quantum critical system lies at the 
boundary ($z = 0$) of the bulk geometry.

The most convenient way to establish the AdS/MERA connection is to
compare the procedures to compute the entanglement entropy in both cases. 
In the classical gravity limit of AdS/CFT, Ryu and Takayanagi (RT) 
derived a celebrated formula yielding the entanglement entropy of
a region $A$ provided that the (boundary) conformal
field theory describing the critical system admits an holographic
gravity dual \cite{ryu}. In the RT approach, the entanglement entropy is
obtained from the computation of a minimal surface in the dual
higher dimensional gravitational geometry (bulk theory); as a
result, the entanglement entropy $S_A$ in a CFT$_D$
is given by the area law relation,
\begin{eqnarray}
S_{A}=\frac{{\rm Area}(\gamma_{A})}{4G^{(D+1)}_N}~,
\label{arealaw}
\end{eqnarray}
where $D$ is the number of spacetime dimensions of the boundary CFT,
$\gamma_{A}$ is the $D$-dimensional static minimal
surface in AdS$_{D+1}$ whose area is given by
${\rm Area}(\gamma_{A})$ and $G^{(D+1)}_{N}$ is the $D+1$ dimensional Newton
constant. In the RT proposal, looking for the minimal surface 
$\gamma_{A}$ separating the degrees of freedom contained in region $A$ from those
contained in the complementary region $B$, amounts to search for the
severest entropy bound on the information hidden in the
AdS$_{D+1}$ region related with $B$. Despite the RT formula has not 
been rigorously proven its validity is supported by very reassuring 
evidence \cite{ryu,strongsub, myers11}. 

In this paper, the interest will be mainly focused in the 
$D=2 =(1 + 1)$ case, for which the equation (\ref{arealaw}) reduces to,
\begin{eqnarray}
S_{A}=\frac{{\rm Length}(\gamma_{A})}{4G^{(3)}_N}~.
\label{RTarealaw}
\end{eqnarray}

Namely, there is a striking similarity between the formula (\ref{RTarealaw}) 
and the computation of the entanglement entropy of a region $A$ with MERA.
The entanglement entropy in MERA is fixed by the structure of the 
\emph{causal cone} \cite{vidalmera, merait}.
The causal cone ${\mathcal {CC}}(A)$ of a region
$A$ of $\ell$ sites, is determined by all the disentanglers and isometries 
along all the levels of the tensor network, which
are directly connected with the sites in the original region $A$,
(see figure 6). As a result, to compute the entropy $S_{A}$ it is 
necessary to trace out any site within the tensor network which does 
not lie in ${\mathcal {CC}}(A)$. The boundary of ${\mathcal {CC}}(A)$ 
is a curve $\widetilde{\gamma}_{A} = \partial\, {\mathcal {CC}}(A)$. 
The length of $\widetilde{\gamma}_{A}$ counts the number of bonds in 
MERA linking the ${\mathcal {CC}}(A)$ with the rest of the sites and, 
when multiplied by the maximal contribution to entanglement entropy given 
by each bond lying in $\widetilde{\gamma}_{A}$, it provides an upper bound 
for the entropy $S_{A}$ \cite{vidalmera, swingle09, Evenbly11},
\begin{eqnarray}\label{entropy:bound}
S_{A} \leq {\rm Length}(\widetilde{\gamma}_{A})\, \log W~,
\end{eqnarray}

where $W$ is the dimension of the auxiliary space of disentanglers 
and isometries. For instance, for disentanglers $\chi_{v_1\, v_2}^{u_1\, u_2}$, 
$u_{1,\, 2} \in \left\lbrace 1, \cdots, W\right\rbrace$ and 
$v_{1,\, 2}\in \left\lbrace 1, \cdots, W\right\rbrace$.

The close connection with the RT formula emerges
as one realizes that $\widetilde{\gamma}_{A}$ can be regarded as
the minimal curve $\gamma_A$ in the RT proposal, since it counts the 
minimal number of bonds which must be considered in order to 
define the bipartition in the MERA tensor network.
Indeed, the minimal curve $\widetilde{\gamma}_{A}$ in an optimized
scale invariant MERA network \cite{Evenbly11} saturates 
the bound given in (\ref{entropy:bound}); this has been
confirmed by explicit computation in one dimensional critical
systems, where it has been shown that \cite{vidalmera},
\begin{eqnarray}\label{mera:EE}
S_{A} \sim k \cdot \log \ell.
\end{eqnarray}

 where $k$ is a constant of order one. Since $\widetilde{\gamma}_{A}$ is defined 
 as the boundary of the ${\mathcal {CC}}(A)$, it
can be interpreted as an holographic screen which optimally
separates the region in the MERA tensor network related with the degrees of
freedom of $A$, from those in its complementary region $B$.

\begin{figure}[t]
  \centering
  \includegraphics[width=0.60\textwidth]{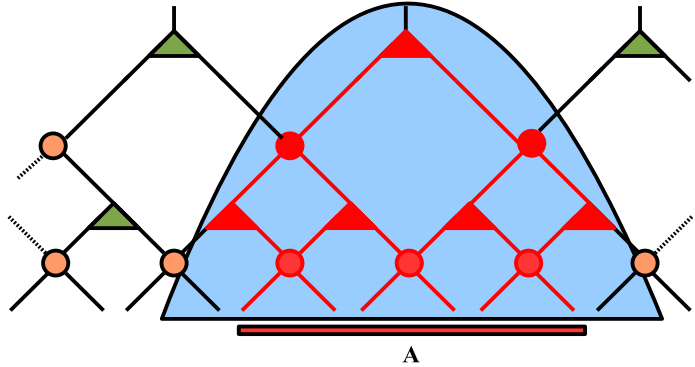}
  \caption{The causal cone of a block of sites in MERA (in red) 
  is given by the minimal number of tensors that may affect 
  to the expectation value of any observable with support on 
  those sites. The boundary of the causal cone of an interval 
  $A$ defines a curve $\widetilde{\gamma}_A$ which length, 
  in an optimized MERA, acts as the bound for the entanglement 
  entropy of the region $A$ given in equation (\ref{entropy:bound}).}
\end{figure}

Despite a complete characterization of the AdS/MERA duality is still lacking, 
very recent advances point to fulfill this objective. 
In \cite{ads_mera_tak}, authors have used a continuos version 
of MERA (cMERA, \cite{cMERA}), the holographic formula for the entanglement 
entropy and the concept of distance between quantum states to address one 
of the major issues concerning the AdS/MERA conjecture, i.e, how 
gravity duals arise from the structure of the quantum entanglement 
of the quantum critical systems under consideration. The question about the role 
of the large $N$ limit in the AdS/MERA duality is also considered in \cite{ads_mera_tak}
as well as in \cite{swingle12}, where it is shown how a number of features of the 
holographic duality in the large $N$ limit emerge 
naturally from entanglement renormalization tensor networks.
 
\section{Holographic geometries for hybrid MERA-MPS networks}\label{MERA_geom:Sec}
It is widely accepted that, non-local observables such 
as two-point functions, entanglement entropy and Wilson 
loops turn out to be crucial for elucidating the subtleties 
of the AdS/CFT correspondence  \cite{adscftbib}. In this section, 
we propose a set of sensible \emph{ans\"atze} geometries which 
provide a consistent behaviour between 
the holographic entanglement entropy (\ref{RTarealaw}) and 
the entanglement entropy computed through the hybrid MERA-MPS tensor 
network presented in section (\ref{Hybrid:Sec}).

The main features of the truncated hybrid MERA-MPS network, are 
the characteristic length scale $\xi$ which fixes the depth 
of the \emph{curtain} and the MPS tensor at the top of the \emph{curtain} 
which accounts for the gapped nature and the non trivial entanglement 
of the IR fixed point. On the other hand, the most salient feature of 
the holographic geometry dual to a system with a characteristic 
length scale $z_0$, is the capping or truncation of the geometry 
for values of the radial coordinate close to this length scale. 
The geometry remains approximately AdS for values of the radial 
coordinate smaller than the characteristic length scale and the 
capping of the geometry implies an infrared (IR) fixed point at 
a finite energy scale ($\sim 1/z_0$). 

\subsection{A geometric ansatz for the hybrid tensor network}\label{ansatz}
In the context of AdS/MERA, it is natural that the \emph{ansatz} 
metric to be associated to the hybrid tensor network corresponds 
to an asymptotically AdS$_3$ spacetime with an IR capped region located 
at $z_0$. In the Poincar\'e patch, this metric, in the constant time slice,
 can be written as, 
\begin{eqnarray}\label{tns_metric}
ds^{2}_{0}= \frac{dz^2}{A(z)^{2}} + B(z)^2\, dx^2~. 
\end{eqnarray}

In these coordinates, the boundary of the spacetime 
lies at $z = 0$ and, requiring that (\ref{tns_metric}) to be 
asymptotically AdS$_3$ amounts to impose that, 
the functions $A(z)$ and $B(z)$ which only depend on the 
radial coordinate $z$, behave as,
\begin{equation}\label{asymp:limits}
A(z)^2 \sim \frac{z^2}{L_{\rm AdS}^2}\, \quad B(z)^2 \sim \frac{L_{\rm AdS}^2}{z^2}~,
\end{equation}

when $z \to 0$. 
The capping of the metric at a finite length 
scale $z_0$ is acomplished through the function $f(z)$,
\begin{eqnarray}\label{ftns_metric}
f(z) = 1 + Q\, \left(\frac{z}{z_0} \right) \log \left(\frac{z}{z_0} \right) - \left(\frac{z}{z_0} \right)^{2}~, 
\end{eqnarray}

 which defines (\ref{tns_metric}) by making,
 \begin{equation}\label{ftns_metric2}
 A(z)^2 = \frac{z^2\, f(z)}{L_{\rm AdS}^2}\, \quad B(z)^2 = L_{\rm AdS}^2\, \frac{f(z)}{z^2}~.
 \end{equation} 
 
 The function $f(z)$ is always positive and does not posses 
 any singularity or zero as long as $0<z<z_0$ and the values of 
 the parameter $Q$ are restricted to lie within $0\leq Q \leq 2$. 
 The asymptotic requirements for (\ref{tns_metric}) are immediately 
 fulfilled by noting that $f(z) \to 1$  when $z \to 0$, while the 
 capping of the geometry follows from $f(z_0)=0$.
 
 As it will be  shown, the function  $f(z)$ controls the growth of the minimal 
 surfaces  appearing in the holographic  computations of entanglement entropy and 
 two point functions near the capped  region of the geometry. In other words,
  different values of the parameter $Q$  give rise to different IR behaviours 
  for the entropy and the correlators. 
   
 As commented in section (\ref{Hybrid:Sec}), different configurations 
of the hybrid MERA-MPS tensor network result in significant distinct 
behaviours of the long range entanglement and the correlations. 
In our approach, once a sensible \emph{ansatz} such as (\ref{tns_metric}) 
is provided, the problem is to show that the behaviour of the entropy and 
correlations in different arrangements of the tensor network are consistent 
with a geometric computation which only depends on $f(z)$ \cite{swingle12}. 

This approach differs from the one posed in \cite{ads_mera_tak}. 
Instead of our consistency-inverse-like problem, authors take a 
more direct approach and use cMERA \cite{cMERA} 
and the concept of distance between quantum states, to compute a  
quantum metric which then is identified with an holographic metric. 
The identification uses the connection between the entanglement entropy 
in MERA and the holographic entanglement entropy commented in section 
(\ref{ads:mera}). In the case of a free scalar theory with mass 
$m$, this procedure yields an spatial part for the associated
 bulk metric given by,

\begin{equation}\label{ryu_metric}
ds^{2}_{0} = \frac{dz^{2}}{4 z^{2}} + \frac{1}{z^2}\left[1 - \left(\frac{z}{z_0}\right)^2 \right]\, dx^{2}~,
\end{equation}

with $z_0\equiv 1/m$. The geometry is capped off at $z_0$, which 
is consistent with the mass gap in the scalar field theory, 
and the vanishing of the metric at $z_0$ is controlled by a 
function which coincides with our \emph{anstaz} (\ref{ftns_metric}) 
for $Q =0$. It is necessary to recall that in the case of such a 
weakly coupled theory, it might be expected that the gravity dual 
(\ref{ryu_metric}) cannot be related to a solution of Einstein gravity 
in three dimensions. Nonetheless, the procedure provides a metric 
which describes the cMERA structure in a qualitative way. 

Finally, it is worth to note that in \cite{espe12}, it has been proposed an 
algorithm to find stationary solutions to 3D Einstein gravity with 
a self-interacting scalar field. The algorithm bootstraps the solutions 
by requiring a geometric input in terms of a function which is close in spirit 
to (\ref{ftns_metric}). The appropiate choices for the input function 
are subjected to severe physical constraints and it would be the matter 
of a future study to use the algorithm to investigate if 
(\ref{ftns_metric}) may be related with some of these solutions.

\subsection{Geometric computation of Entanglement Entropy and Correlators}\label{holo_comp:sec} 
 In order to obtain the entanglement entropy of an static 
 region $A$ lying at the boundary of an asymptotically AdS 
 spacetime, the Ryu and Takayanagi proposal \cite{ryu} 
 accomplishes the task throgh the computation of an static 
 surface of minimal area in the bulk geometry. In AdS$_3$, 
 this means a static curve of minimal length $\gamma$, i.e 
 a geodesic.  
Thus, let us consider the region $A$ in the boundary of 
(\ref{tns_metric}) given by an interval with length $\ell$ 
along $x$ direction.
 Choosing the origin in the center of this interval, the 
 symmetry of the problem allows us to restrict to curves 
 described by an even function $z = z(x)$ embedded in the 
 constant time slice of (\ref{tns_metric}). The metric 
 induced on those curves reads
\begin{equation}\label{induced} 
ds^2_{\rm ind}=\left(\frac{\dot{z}^2}{A(z)^2} + B(z)^2\right)\, dx^2~,
\end{equation}

where $\dot{z}=dz/dx$. To compute the length of the minimal 
curve $\gamma$ amounts to integrate $\sqrt{{\rm det\, (g_{ind})}}$ 
over the interval $A$ i.e,
\begin{equation}\label{area_def}
{\rm Length}(\gamma)=2\, \int_0^{\ell/2}\, \sqrt{{\rm det\, (g_{ind})}}\, dx = 2\, \int_0^{\ell/2}\, dx\,  \sqrt{\frac{\dot{z}^2}{A(z)^2} + B(z)^2}~,
\end{equation}

where ${\rm g_{ind}}$ is the induced metric 
on the static curve $\gamma$.

 Considering the integrand of (\ref{area_def}) as 
 a Lagrangian density $\mathcal{L}[\dot{z}, z]$, 
 one notices that it does not explicitly depends on $x$. 
 Namely, the independence of $\mathcal{L}[\dot{z}, z]$ 
 on $x$ leads to the conserved quantity 
 $\mathcal{H}= p_z\, \dot{z} - \mathcal{L}$, 
 where $p_z = \partial{\mathcal{L}}/\partial{\dot{z}}$. 
 In particular, one gets 
 
 \begin{equation}\label{hamilt:eq}
 \mathcal{H} = -\frac{B(z)}{\sqrt{1 + (\dot{z}^2/A(z)^2\, B(z)^2)}}~. 
 \end{equation}

 Since at the maximum value of the radial coordinate $z_{\rm max}$, 
 we have $\dot{z}=0$, the constancy of $\mathcal{H}$ can 
 be expressed by settling that $\mathcal{H}^2 = B(z_{\rm max})^2$ 
 which allow us to write (\ref{area_def}) as, 
 \begin{equation}\label{geod:length}
 {\rm Length}(\gamma)=-2\, \int_{z_{\rm max}}^{\epsilon}\, \frac{B(z)\, dz}{A(z)\, \sqrt{B(z)^2 - B(z_{\rm max})^2}}~,
 \end{equation}

where $\epsilon$ is an UV-cutoff such that the 
integration domain of the radial coordinate constricts
to $(\epsilon, z_{\rm max})$. Furthermore, it is worth 
to express the relation between the length of the interval 
$A$ as a function of $z_{\rm max}$ given by,
\begin{equation}\label{length_max}
\ell_{\rm max}\equiv \ell(z_{\rm max})=-2\, \int_{z_{\rm max}}^{\epsilon}\, \frac{B(z_{\rm max})\, dz}{A(z)\, B(z)\, \sqrt{B(z)^2 - B(z_{\rm max})^2}}~.
\end{equation}
	
The equation (\ref{geod:length}) allow us to compute 
the entanglement entropy of the boundary interval $A$ 
using the RT formula (\ref{RTarealaw}). In addition, 
the two point functions of scaling  operators $\Phi$ 
inserted at two spacelike separated  points on the 
boundary of AdS are also considered. 
 According to \cite{adscftbib}, this correlator 
 $\mathfrak{C}^{[\mathcal{D}]}_{\Phi}$, is obtained 
 from the spacetime propagator $\mathfrak{C}^{[\mathcal{D}]}_{\rm Holo}$ 
 between the corresponding points on the UV-cutoff 
 boundary lying at $z \sim \epsilon$ by simply writting, 
\begin{eqnarray}\label{twopoint:corr}
\mathfrak{C}^{[\mathcal{D}]}_{\Phi} = \epsilon^{-2\Delta} \, \mathfrak{C}^{[\mathcal{D}]}_{\rm Holo}~,
\end{eqnarray}
 
where $\mathcal{D}$ is the distance between the inserted 
operators on the boundary and $\Delta$ is the scaling 
dimension of the operator $\Phi$. The holographic 
propagator $\mathfrak{C}^{[\mathcal{D}]}_{\rm Holo}$, 
in the leading order of its semi-classical approximation, 
is given by a sum over geodesics $\Upsilon_{\mathcal{D}}$ 
connecting the points on the cut-off boundary,
\begin{eqnarray}\label{geod:propagator}
\mathfrak{C}^{[\mathcal{D}]}_{\rm Holo} = \sum_{\Upsilon_{\mathcal{D}}}\, \exp \left[-m\, {\rm Length}(\Upsilon_{\mathcal{D}})\right] \sim \exp \left[ -m\, {\rm Length}(\gamma_{\mathcal{D}})\right] ~.
\end{eqnarray}

Here, ${\rm Length}(\gamma_{\mathcal{D}})$ represents the length 
of the shortest geodesic connecting the two boundary 
points and $m$ is the mass of the fields $\phi$ dual 
to the operators $\Phi$. Namely, in the AdS/CFT 
correspondence, the scaling dimensions of  boundary 
operators are related to the masses 
$m$ of their dual fields. For instance, 
when $\Phi$ is an scalar operator of large scaling dimension 
$\Delta$, then it holds that $\Delta \approx m\, L_{AdS}$ \cite{adscftbib}.

\subsection{Entanglement and Correlators in the ansatz geometry}\label{holoresults}
In this subsection, we use the generic holographic formulas 
(\ref{geod:length}) and (\ref{length_max}) particularized for 
(\ref{ftns_metric}) and (\ref{ftns_metric2}). A simple substitution 
yields,
\begin{equation}\label{geod:calc}
{\rm Length}(\gamma)=-2\, L_{\rm AdS} \int_{z_{\rm max}}^{\epsilon}\, \frac{z_{\rm max}\, dz}{z\, \sqrt{z_{\rm max}^2\, f(z) - z^2\, f(z_{\rm max})}}~,
\end{equation}

and
\begin{equation}\label{length:calc}
\ell_{\rm max}=-2\, \sqrt{f(z_{\rm max})}\, \int_{z_{\rm max}}^{\epsilon}\, \frac{z\, dz}{f(z)\, \sqrt{z_{\rm max}^2\, f(z) - z^2\, f(z_{\rm max})}}~.
\end{equation}

Our main interest here, lies in characterizing the entanglement 
entropy contributions given by the IR regions of the geometry 
\cite{soliton, barbon, klebanov}. 
Thus, we are mainly concerned with those regimes for which 
$z_{\rm max} = z_0$. When $z_{\rm max} \ll z_0$, the values 
of the radial coordinate $z$ keep close to the AdS boundary and 
$f(z)\to 1$. For these regimes one gets,
\begin{equation}\label{conformal}
{\rm Length}(\gamma)\approx 2\,  L_{\rm AdS} \, \log \frac{\ell_{\rm max}}{\epsilon}~, \quad \ell_{\rm max} \approx 2\, z_{\rm max}~,
\end{equation}

i.e, one obtains the scaling of the entanglement entropy in 
a pure AdS spacetime, which is something to be expected as far as 
the geometry (\ref{tns_metric}) reduces to pure AdS when $z \ll z_0$. 

 Contrarily, when $z \to z_0$ the function $f(z)$ can be written as,
\begin{equation}\label{nearhoriz}
f(z)\vert_{z \to z_0} \approx \frac{1}{z_0^2}\, (z_0^2 - Q\, z_0\, z + (Q-1)\, z^2)~.
\end{equation}

As $z_{\rm max}\equiv z_0$ and $f(z_0)=0$, the geodesic length and $\ell_{\rm max}$
read as,
\begin{equation}\label{geod:nearhor}
{\rm Length}(\gamma) = -2\,  L_{\rm AdS} \, \int_{z_0}^{\epsilon}\, \frac{z_0\, dz}{z\, \sqrt{z_0^2 - Q\, z_0\, z + (Q-1)\, z^2}} =
2\,  L_{\rm AdS} \, \log \frac{\ell_{\rm max}(\delta)}{\epsilon}~, 
\end{equation}

and
\begin{equation}\label{lmax}
\ell_{\rm max}(\delta) = -2\, \sqrt{h(z_0)}\, z_0\, \int_{z_0}^{\epsilon}\, \frac{z\, dz}{(z_0^2 - Q\, z_0\, z + (Q-1)\, z^2)^{3/2}}=\frac{4\, z_0}{\delta}~,
\end{equation}

with $\delta = 2 - Q$ and $h(z)=z_0^2 - Q\, z_0\, z + (Q-1)\, z^2$. 

 These results imply that the geodesic length in
the \emph{ans\"atze} geometries parametrized by $\delta$, 
saturate when considering intervals $A$ whose length 
$\ell > \ell_{\rm max}(\delta)=4\, z_0/\delta$. 

 For the sake of forthcoming discussions, let us illustrate the 
behaviour of the geodesic curves by simply recasting  
their length as, 
\begin{equation}\label{geod:nearhor2}
{\rm Length}(\gamma) = 2\,  L_{\rm AdS} \, \left(\log \frac{2\, z_0}{\epsilon}  + \log \frac{2}{\delta}\right)~.
\end{equation}

The first term in (\ref{geod:nearhor2}) amounts to the 
geodesic distance between the boundary points of $A$
located at $z \sim \epsilon$ and the capping region lying 
at $z_0$. In other words, this term is related with the piece 
 of the curve connecting the boundary of the geometry and the IR capping 
region. On the other hand, the second term, may be visualized 
as follows: the geodesics, once arriving at the capping region, 
instead of finishing, stroll over the IR region to a maximum 
allowed extent quantified by this term. 

 The geodesic length (\ref{geod:nearhor2}) thus 
 permit us to identify two different contributions in 
 the holographic entanglement entropy. Indeed, by 
 inserting  (\ref{geod:nearhor2}) into (\ref{RTarealaw}), 
 one gets,
\begin{equation}\label{contributions}
S_A = S_{A}^{\, \rm UV} + S_{A}^{\, \rm IR}(\delta)~, \quad S_{A}^{\, \rm UV}\propto \log \frac{2\, z_0}{\epsilon}~, \quad S_{A}^{\, \rm IR}(\delta)\propto \log \frac{2}{\delta}~.
\end{equation}

The term $S_{A}^{\, \rm IR}(\delta)$ only depends on the parameter 
which characterizes the geometry at the capping region and its 
behaviour is quite rich within the range $0 < \delta \leq 2$ 
(i.e, $0 \leq Q < 2$); while it remains finite for $0 < \delta < 2$,
vanishes for $\delta=2$ and it diverges when $\delta \to 0$.

As the geodesic length is
saturated for intervals $A$ with 
$\ell > \ell_{\max}$, then, according to (\ref{twopoint:corr}, 
\ref{geod:propagator}), the two point correlators of 
operators inserted at the boundary of the geometry also show 
a similar saturation phenomena as shown by,
\begin{eqnarray}\label{correlator_extr}
\mathfrak{C}^{[\ell]}_{\rm Holog} =\epsilon^{-2\Delta}\, \exp \left[ -2 \Delta\, \log \frac{\ell_{\rm max}(\delta)}{\epsilon}\right] =
\ell_{\rm max}(\delta)^{-2 \Delta}~. 
\end{eqnarray}

This result implies that in a capped geometry defined by $\delta$, 
all the correlators vanish when the points at the boundary are separated by 
a distance $\ell > \ell_{\rm max}(\delta)$. Namely, it is convenient 
to interpret $\ell_{\rm max}$ as an \emph{effective} correlation length,
\begin{equation}\label{hol:xi}
\zeta_{\rm Holo}=\ell_{\rm max}(\delta)~.
\end{equation}

This effective corelation length diverges as $\delta \to 0$ while is 
minimal for $\delta=2$, i.e ${\rm min} \left\lbrace \zeta_{\rm Holo}\right\rbrace  = \ell_{\rm max}(\delta=2)= 2\, z_0$.

\subsection{Entanglement in MERA-MPS networks and geometry}\label{EEgeom:sec}
In this subsection, we show how the features of 
the long range entanglement in an hybrid MERA-MPS 
state are consistently represented by the ansatz geometry 
(\ref{tns_metric}). With this aim, we consider a tensor network 
composed of a MERA \emph{curtain} with $\tau_0 = \log \xi$ layers
so that $\xi = 2\, z_0$. In addition, on the top of the 
\emph{curtain}, we place an MPS tensor state $\ket{\mathcal{C}_{\rm MPS}}$ 
with bond dimension $\mathcal{W}$.

Let us first to compute the entanglement entropy of a 
region $A$ with $\ell \gg \xi$. In the MERA \emph{curtain}, 
the causal cone of the region $A$ is connected to the rest 
of the MERA sites through a number $n(A)$ of traced out bonds 
which is proportional to $\log \xi$ and not depends on $\ell$. 
This yields a saturated value for the contribution of the MERA 
layers to the entanglement entropy of the block $A$ given by,
\begin{equation}\label{entropy:curtain}
S_A^{\, \rm MERA}\propto  \log \xi = \log 2\, z_0~.
\end{equation} 

The second contribution to the entanglement entropy 
of $A$ comes from the top MPS tensor. This IR state provides 
an additional contribution $S_A^{\, \rm MPS}$ which is bounded by
$\log \mathcal{W}$ (see eq.(\ref{EE:mps})). Thus, the total 
entanglement entropy amounts to,
\begin{eqnarray}
S_A = S_A^{\, \rm MERA} + S_A^{\, \rm MPS} \leq \log \xi + 2\, \log \mathcal{W}~.
\end{eqnarray}

Attending to (\ref{contributions}) one realizes that: first, the term $S_A^{\, \rm UV}$ 
is equal to the contribution given by the MERA \emph{curtain}. On the other hand, 
the MPS contribution to the entanglement entropy may be interpreted in terms of 
the associated holographic geometry if one identifies this contribution with 
$S_A^{\, \rm IR}$  in (\ref{contributions}). This relates the  parameter 
$\delta$ managing the growth of the geodesic length in the capping region of 
the geometry (\ref{tns_metric}),  with the bond dimension 
$\mathcal{W}$ which characterizes the long range entanglement 
in the tensor network, i.e,
\begin{equation}\label{ent:geom1}
\mathcal{W} \geq \frac{2}{\delta}~.
\end{equation}
 
The last expression allow us to consistently relate 
different configurations of the tensor network  with  
different values of the geometric parameter $\delta$.
As an example, if the top tensor is a trivial non 
entangled product state with bond dimension $\mathcal{W}=1$, 
the contribution to the entropy it yields is identically zero. 
This patten of long range entanglement is geometrically 
represented by taking $\delta = 2$. Conversely, one might say 
that there is no tensor network representation when taking 
$\delta \to 0$ as this requires a $\mathcal{W}\to \infty$.

The next thing we would like to consider is the two point function 
of an scaling operator $\Phi$ inserted at two sites in the 
bottom of the MERA \emph{curtain} and separated by a distance $\ell \gg \xi$. 
Remarkably, the analysis of the two point functions also provides a 
significative  connection between the IR entanglement structure  
of the tensor network and the \emph{ansatz} geometry (\ref{tns_metric}). 
Assuming the translational invariance of the system, this correlator can be 
written as 
\begin{equation}\label{meracorr:sec44}
\mathfrak{C}^{\ell}_{\rm TNS} = \bra{\Psi} \Phi(-\ell/2)\,  \Phi(\ell/2)\ket{\Psi} ~, 
\end{equation}

with $\ket{\Psi}$ being the tensor decomposition of the system given 
by the hybrid MERA-MPS network. To perform the computation, one 
needs to carry out $\tau_0 = \log \xi$ contractions in the MERA 
\emph{curtain} and then an expectation value with the top 
MPS state  $\ket{\mathcal{C}_{\, \rm MPS}}$. Noticing that 
$\Delta\equiv -\log \mu$, and (\ref{MPS:2point}) one obtains,
\begin{eqnarray}\label{tns:corr44}
\mathfrak{C}^{\ell}_{\rm TNS} = 
\xi^{-2 \Delta}\, \bra{\mathcal{C}_{\, \rm MPS}}\Phi(-\ell/2\, \xi)\,  \Phi(\ell/2\, \xi)\ket{\mathcal{C}_{\, \rm MPS}} =
\xi^{-2 \Delta}\, \sum_{\vert\lambda_{\nu}\vert < 1} c_{\nu}\, (\lambda_{\nu})^{\ell/\xi}~.
\end{eqnarray}

Here, it is reasonable to assume that the leading term 
controlling the decay of the correlator (\ref{tns:corr44}) 
is given by the biggest $\lambda_{\nu} < 1$, i.e $\lambda_2$. 
Furthermore, the leading decaying term can be expressed as an 
exponential  with an \emph{effective} correlation length given by,
 \begin{equation}\label{mps:corrlength}
  \zeta_{\, \rm TNS}=-\frac{1}{\log \vert \lambda_{\Gamma}\vert}~, \quad \lambda_\Gamma = \lambda_2^{1/\xi} \quad \Longrightarrow \quad  \zeta_{\, \rm TNS}=-\frac{2\, z_0}{\log \vert \lambda_{2}\vert}~.
  \end{equation}

Regarding the formal equivalence between the 
entanglement entropy in the MERA-MPS network and 
the holographic entanglement entropy in the ansatz 
geometry (\ref{tns_metric}) it is reasonable to 
connect their \emph{effective} correlation 
lengths. The simplest assumption is to establish 
an equivalence up to a 
constant term $\kappa$ so that,
\begin{eqnarray}\label{equiv:xi}
\zeta_{\rm Holo}=\frac{4\, z_0}{\delta}&=&\kappa\, \zeta_{\, \rm TNS}=-\frac{2\, \kappa\, z_0}{\log \vert \lambda_{2}\vert}~,\nonumber \\ 
\lambda_2 &=& \exp \left( -\frac{\kappa}{2}\, \delta\right)~. 
\end{eqnarray}

These expressions relate the piece of 
$\operatorname{Spec}\left( \mathbb{E}_{\mathbf{1}}\right) $ 
controlling the leading behaviour in 
the decay of correlators, with the \emph{holographic} 
parameter $\delta$ in a consistent way. Ideed, if 
$\delta \to 0$ then $\lambda_2 \to 1$, reflecting 
the divergence in $\zeta_{\, \rm TNS}$ which one might 
expect for an MPS with $\mathcal{W} \to \infty$. 
On the other hand, when $\delta=2$, $\lambda_2$ can be 
made parametrically small if $\kappa \sim 1$, i.e 
 $\lambda_2 = e^{-\kappa}$. This is consistent with an 
MPS with a bond dimension $\mathcal{W}\sim 1$, i.e close 
to be a non entangled product state.

\subsection{Mapping MERA to MPS networks}

In this subsection we analize the decomposition of the geodesic 
length (\ref{geod:nearhor}) into the two terms shown in (\ref{geod:nearhor2}) 
from a tensor network perspective. With this aim, we note that
the holographic entanglement entropy in terms of (\ref{geod:nearhor}) 
given by,

\begin{equation}\label{hee}
S_A \propto \log\, \frac{\ell_{\rm max}(\delta)}{\epsilon}~,
\end{equation}

may be generically represented  by a tensor network composed 
only by a finite number of MERA layers, $\widehat{\tau} = \log \widehat{\xi}(\delta)$ 
so that,

\begin{equation}\label{gen:mera}
\widehat{\xi}(\delta)=\ell_{\rm max}(\delta)=\frac{4\, z_0}{\delta}~.
\end{equation}

Nevertheless, as being concerned with 1-D gapped phases possesing a non 
trivial exponential decay of correlations for distances 
$\ell \gg 2\, z_0$, it is thus important to note here that,  
for a finite range MERA network, is not possible, in general, 
to yield such a behaviour. Namely, only a non homogeneous finite 
range MERA with $\widehat{\tau}$ layers can match (\ref{hee}) 
while implementing exponentially decaying correlations 
for $\ell \gg 2\, z_0$.  This network is composed by a 
\emph{curtain} of $\tau_0 = \log 2\, z_0$ layers with identical 
disentanglers and isometries and 
a fixed number $\Delta\, \tau= \log\, 2/\, \delta$ of top layers  
where the tensors are allowed to be different from those composing the 
\emph{curtain} \cite{Evenbly11}. In other words, the layered structure of the
non homogeneous finite range MERA can be naturally 
decomposed as,
\begin{eqnarray}\label{mera:dec}
\widehat{\tau}&=& \tau_0 + \Delta\, \tau~, \nonumber \\
\widehat{\tau}&=& \log 2\, z_0 + \log \frac{2}{\delta}~.
\end{eqnarray}

This is the tensor network justification for considering the 
arbitrary decomposition of the geodesic length given in 
(\ref{geod:nearhor2}). However, it is still difficult to 
explicitly show how this MERA network implements exponentially 
decaying correlations at long distances. To this end it is 
worth to recall that for a 1-D finite range MERA made of 
$\Delta\, \tau$ layers of tensors with a bond dimension $W^{*}$, 
there is map by which it may be re-expressed 
as an MPS with bond dimension $\mathcal{W}$. This map read as,
\begin{equation}\label{map}
\mathcal{W} = (W^{*})^{\Delta\, \tau}~.
\end{equation}

Assuming that $\log (W^{*}) \sim \mathcal{O}(1)$, 
if one applies this map to the $\Delta\, \tau$ top layers 
of the inhomogeneous finite range MERA, one gets 
the hybrid MPS-MERA network of section \ref{EEgeom:sec}
with $\mathcal{W} \sim 2/\delta$ (Eq. \ref{ent:geom1})
and the ability to use the analytical bonanza for exponentially 
decaying correlations offered by MPS which also lead to establish 
the relationship (\ref{equiv:xi}). 

\section{Discussion and Concluding Remarks}
 By exploring some hints posed in \cite{soliton}, 
 new insights on the connection between the structure 
 of MERA states and their potential geometric descriptions 
 have been provided. The striking relationship between 
 the MERA and AdS/CFT, initially posed in \cite{swingle09}, 
 manifests through the use of the holographic entanglement 
 entropy \cite{ryu}. Both proposals establish a connection 
 between entanglement and geometry. While in the former, 
 one tries to ascribe a geometry to a known entanglement 
 structure in the tensor network, in the latter, it is 
 the behaviour of the entanglement which comes determined by 
 the known dual geometry of the boundary quantum system. 

 Here, we have shown that the entanglement entropy and the 
 two point functions computed through a tensor network 
 state representing a one dimensional gapped quantum many 
 body system, remains consistent, at least at the qualitative 
 level, with a geometric computation of both quantities once 
 a sensible metric associated to the tensor network is provided. 
 The metric corresponds to an asymptotically AdS geometry with 
 an infrared capping region parameterized in terms of a function. 
 This function characterizes the growth of minimal surfaces near 
 the capped region of the geometry and the holographic computation 
 depends only on it. As different arrangements of the network 
 provide significant distinct behaviours of the entropy and 
 correlations, the  \emph{ans\"atze} metrics are proposed in a 
 kind of inverse-like problem in order to obtain some qualitative 
 matches between both computations. 

From these matches, we have presented some expressions relating 
the entanglement structure of the tensor network and the parametrizing 
function of the associated geometry. Indeed, the equation (\ref{equiv:xi}) 
relates the spectrum of the tranfer matrices of the top MPS tensor with 
the parameter $Q$ defining the IR behaviour of the geometry. This kind 
of result may be useful in order to, \emph{via} AdS/TNS, open the 
possibility to classify the gapped phases of 1D-quantum many-body 
systems in terms of their associated geometries. As posed in 
\cite{wen11, davidpe11, pollman12}, different phases (including 
symmetry protected topological phases) of 1D gapped systems with 
an MPS representation, can be characterized by the spectrum of 
the MPS tranfer matrices. 

It is necessary to recall that, despite the consistency requeriments 
commented above, the \emph{ans\"atze} metrics cannot be related 
to any solution of Einstein gravity. Indeed, our approach only provides 
a metric which describes the entanglement structure in the tensor network 
in a qualitative way. In this sense, in \cite{espe12}, it has been 
proposed an algorithm to find stationary solutions of the three dimensional 
Einstein gravity with an interacting scalar field. The algorithm 
bootstraps the solutions by requiring a geometric input in terms of 
function which is close in spirit to our parametrizing function. 
It would be the matter of a future study to use similar ideas 
in order to investigate if our \emph{ans\"atze} may be related with 
solutions of a sensible gravitational theory.

  Finally, the proposal of this paper could be extendable to higher 
  dimensions. Namely, there are 2D generalizations of MERA \cite{mera2} 
  and  MPS (PEPS) \cite{Verstraete09} to reasonably construct 2D 
  versions of   the hybrid network proposed here. For instance, in 
  \cite{meratoric},   authors presented the MERA tensor network for 
  the Kitaev's toric code model \cite{kitaev:tc}, the simplest 
  topologically ordered phase in two spatial dimensions. The MERA tensor 
  network for the toric code model is capped off by a top tensor which 
  encodes the information about the topological order. Furthermore, 
  and in a closer analogy to the tensor networks considered here, 
  the toric code also possess a very simple representation in terms of 
  a PEPS with a bond dimension $\mathcal{W}=2$ \cite{peps}. Indeed, for 
  any PEPS with a finite bond dimension, it is always simple to calculate 
  the topological entanglement entropy defined in \cite{kitaev:top}, which 
  is one of the major markers of topological order in gapped systems.
  
  In addition to the toric code, there are more complex (2+1) 
  topological phases which admit a tensor network description such 
  as the Levin-Wen (LW) string-net states \cite{stringnet, tensorstrings, 
  biamonte12}. String-net states are lattice models representing a large 
  class of \emph{doubled} topological phases i.e, topological phases with 
  time reversal symmetry preserved. The ground state of a LW state can 
  be viewed as the fixed point state of some renormalization group flow 
  and it is conjectured that its properties can be described by a 
  \emph{doubled} Chern-Simons theory. Recently, a supersymmetric version 
  of a doubled Chern-Simons theory has attracted a strong attention in 
  string theory \cite{abjm}.   The model is a CS theory with a 
  ${\rm U(N)}_k \times {\rm U(N)}_{-k}$ gauge group in addition to 
  bifundamental matter fields (scalars and fermions) of ${\rm U(N)}$. 
  These theories have been argued to be dual to M-theory on 
  ${\rm AdS}_4 \times {\rm S^7}/\mathbb{Z}_k$. 
  In this context, it is thus interesting to seek for ansatz geometries that 
  qualitatively describe the entanglement structure of the tensor 
  networks associated to doubled topological phases.

 	An equally challenging proposal runs conversely. In \cite{cs:duals}, 
 	authors provide the gravity dual  of a (2+1) dimensional field theory 
 	which in its IR limit flows to a  Chern-Simons theory. Since the theory 
 	is gapped, the geometry is capped  off at IR. To complete the model, 
 	D7-branes are located lying on the  capping region of the geometry, 
 	making the IR behavior different from  that of the "naked" geometry. 
 	 Concretly, the D7-branes are the source  of a Chern-Simons term in 
 	 the action, which gives rise to a topological entanglement entropy 
 	 term through (\ref{arealaw}). It is thus tempting  to apply the ideas 
 	 presented here in order to find 2D tensor network states with top 
 	 tensors whose  entanglement properties match the features observed 
 	 in the geometric duals of \cite{cs:duals}.

 \acknowledgments{JMV is grateful to P. Sodano, J. Prior, S. Montangero, 
 L. Tagliacozzo and G. Sierra for very fruitful insights at different 
stages of this project. JMV has been supported by the Spanish Office for
Science FIS2009-13483-C02-02 and Fundaci\'on S\'eneca Regi\'on de
Murcia 11920/PI/09. I thank the Centro de Ciencias de Benasque Pedro Pascual
 (Spain) and the International Institute of Physics in Natal (Brazil)
for their hospitality at several stages of this project.}

\bibliographystyle{JHEP}

\end{document}